\documentclass{article}
\usepackage{amsmath}

\usepackage{xcolor}
\pagecolor{black}
\color{white}

\title{Willchain: Decentralized, Privacy-Preserving, Self-Executing, Digital Wills}
\author{Jovonni L. Pharr \\ info@willchain.org}
\date{
v0.0.1 May 26, 2023 
\\
v0.0.2 April 1, 2024
}

\begin{document}

\maketitle

\begin{abstract}

This work presents a novel decentralized protocol for digital estate planning that integrates advances distributed computing, and cryptography. The original proof-of-concept was constructed using purely solidity contracts. Since then, we have enhanced the implementation into a layer-1 protocol that uses modern interchain communication to connect several heterogeneous chain types. A key contribution of this research is the implementation of several modern cryptographic primitives to support various forms of claims for information validation. These primitives introduce an unmatched level of privacy to the process of digital inheritance. We also demonstrate on a set of heterogeneous smart contracts, following the same spec, on each chain to serve as entry points, gateways, or bridge contracts that are invoked via a path from the will module on our protocol, to the contract. This ensures a fair and secure distribution of digital assets in accordance with the wishes of the decedent without the requirement of moving their funds. This research further extends its innovations with a user interaction model, featuring a check-in system and account abstraction process, which enhances flexibility and user-friendliness without compromising on security. By developing a dedicated permissionless blockchain that is secured by a network of validators, and interchain relayers, the proposed protocol signifies a transformation in the digital estate planning industry and illustrates the potential of blockchain technology in revolutionizing traditional legal and personal spheres. Implementing a cryptoeconomic network at the core of inheritance planning allows for unique incentive compatible economic mechanisms to be constructed. This yields a unique opportunity to include traditional stakeholders in the inheritance and estate planning industry to become fair contributors, and stakeholders in this form of modernization for their industry. We hope that with this protocol, traditional firms, and service providers in the industry can move towards standardized specification for digital inheritance that remains borderless, open, transparent, and censorship-resistent.

\end{abstract}

\clearpage

\tableofcontents


\section{Current Landscape}
Blockchain Wills \cite{blockchain-wills}, and Crypto Wills \cite{crypto-wills} have been proposed before this work, and have mainly focused on the contract layer. In an ever-changing blockchain ecosystem, these primitives should account for an interchain world, with increasing privacy concerns, and programmability. We expand on the work of our predecessors in this problem space.

\subsection{The Problems with Traditional Wills}

Let us consider $A_{\text{traditional}}$ to denote the set of all assets traditionally considered in a will. With the advent of digital assets, we now have a new set of assets, $A_{\text{digital}}$, which represents a rapidly growing portion of an individual's estate. The union of these sets, $A_{\text{total}}$, should ideally be the scope of a well-planned will, as per equation \ref{eq:asset_union}.

\begin{equation}
A_{\text{total}} = A_{\text{traditional}} \cup A_{\text{digital}}
\label{eq:asset_union}
\end{equation}

However, traditional wills, $W_{\text{traditional}}$, currently only accommodate $A_{\text{traditional}}$, leaving $A_{\text{digital}}$ largely unaccounted for. This discrepancy results in a significant portion of an individual's assets not being distributed as per their wishes. Several circumstances of lost wealth due to death have been documented \cite{blockchain-wills}, and they serve as a reminder that without the proper setup, having no reliable inheritance plan for virtual assets can lead to trapped wealth.

Moreover, the physical nature of traditional wills introduces complications related to storage, access, and authenticity. Let $\alpha$ represent the integrity of a will (measured by its completeness and accuracy), and $\tau$ denote the time elapsed since the will's creation. A function $f(\alpha, \tau)$ could be defined such that it decreases over time, and the probability of a dispute arising, $P_{\text{dispute}}$, is inversely proportional to $f(\alpha, \tau)$.

At the time of writing, there is a trend among current virtual asset inheritance providers whereby they are using multisig accounts, typically 2-of-3. For some providers, they sometimes have centralized flows that allow a beneficiary to submit a claim, whereby the owner can cancel the claim within a time frame. While these are steps in the right direction, there are several inherent challenges with these traditional \(\texttt{Cryptowill}\) approaches.

Transitioning from traditional wills to a digital will system, denoted as $W_{\text{digital}}$, presents its own set of challenges. Ensuring privacy and security in $W_{\text{digital}}$ is paramount. Let $S_{\text{priv}}$ represent the privacy security of a system, where a higher value of $S_{\text{priv}}$ indicates better privacy. The goal is to maximize $S_{\text{priv}}$ while maintaining user-friendliness and affordability.

In the industry, service providers often employ multisig accounts, which, while enhancing security, introduce potential centralization and custody issues. Such models, reliant on external entities for key management, risk compromising user autonomy and privacy. Moreover, they often come with a cost—both financial and operational—that could deter widespread adoption.

From a business perspective, let $C_{\text{develop}}$ represent the cost of developing a secure and user-friendly platform and $C_{\text{user}}$ represent the cost for a user to use the platform. The challenge lies in minimizing $C_{\text{develop}}$ and $C_{\text{user}}$ while maintaining a high $S_{\text{priv}}$ and ensuring full control remains with the user.

\subsection{Technological and Business Challenges}

Transitioning from traditional wills to a digital will system, denoted as $W_{\text{digital}}$, presents its own set of challenges. Ensuring privacy and security in $W_{\text{digital}}$ is paramount. Let $S_{\text{priv}}$ represent the privacy security of a system, where a higher value of $S_{\text{priv}}$ indicates better privacy. The goal is to maximize $S_{\text{priv}}$ while maintaining user-friendliness and affordability.

From a business perspective, let $C_{\text{develop}}$ represent the cost of developing a secure and user-friendly platform and $C_{\text{user}}$ represent the cost for a user to use the platform. The challenge lies in minimizing $C_{\text{develop}}$ and $C_{\text{user}}$ while maintaining a high $S_{\text{priv}}$.

\subsection{Regulatory Landscape}

The legal framework around digital wills, denoted by $L_{\text{digital}}$, varies significantly across different jurisdictions. While some jurisdictions ($J_{\text{pro}}$) have progressive laws regarding digital wills, others ($J_{\text{con}}$) continue to insist on traditional norms. The challenge lies in designing $W_{\text{digital}}$ such that it is valid across all jurisdictions, as represented in equation \ref{eq:legal_challenge}.

\begin{equation}
W_{\text{digital}} \subseteq L_{\text{digital}}(J_{\text{pro}} \cup J_{\text{con}})
\label{eq:legal_challenge}
\end{equation}

In this paper, a solution that addresses these mathematical and practical challenges is proposed. By leveraging the power of blockchain technology and innovative cryptographic techniques, it ensures secure, private, and transparent management of $A_{\text{total}}$. The design ensures its accessibility to individuals with varying degrees of technological proficiency and operability across different jurisdictions, thereby catering to a broad user base. This new paradigm in estate planning mitigates the disadvantages of traditional wills and overcomes the complexities in current digital will platforms. By meticulously integrating legal considerations with robust technology, this approach ensures compliance with regulatory landscapes, making the digital will universally applicable.

Moreover, it reduces the financial overhead of estate planning, making it an affordable solution that does not compromise on the quality or security of service. This novel framework is built with an understanding of the importance of a will - a significant document embodying an individual's last wishes. The rigorous mathematical foundations of this system provide a reliable, efficient, and user-friendly solution, taking a substantial step forward in modernizing estate planning. 

While this paper presents a significant advance in the field of digital wills, it also opens avenues for future research and developments. With the continued digitization of assets and expansion of the digital world, the techniques and methodologies presented here can be further refined and extended. As laws and regulations evolve to keep pace with these advancements, so too must the solutions that we propose and develop. This paper thus represents not only a culmination of present research but also a starting point for future innovations in the realm of digital estate planning.

\section{Traditional Providers Today}
Traditional service providers often navigate this complex landscape by adopting a one-size-fits-all approach, which might not cater to the unique legal and regulatory nuances of each jurisdiction. Our protocol, on the other hand, introduces a modular design that allows for customizable components to comply with specific legal requirements, thereby broadening its applicability and ensuring compliance in a diverse regulatory landscape.

\subsection{Innovative Solutions to Existing Challenges}

To address the challenges identified with traditional service providers, our protocol introduces several key capabilities:

\begin{itemize}
    \item \textbf{Noncustodial}: By employing blockchain technology, we ensure that all digital assets are managed without requiring a centralized authority for key safekeeping. This not only enhances security but also ensures that the user retains full control over their assets.
    
    \item \textbf{Enhanced Privacy and Security}: Leveraging advanced cryptographic techniques, our protocol guarantees a high level of privacy, and anonymity if the will creatr
    
    \item \textbf{Universal Accessibility}: Designed to be fully digital and operable across borders, our protocol ensures that anyone with internet access can manage and execute their digital will, reflecting the true spirit of decentralization inherent in blockchain technology.
    
    \item \textbf{User Control and Flexibility}: Our protocol allows users to define conditions for asset distribution, including but not limited to pre-determined dates. These conditions can be configured at the protocol level, and enhance functionality via smart contracts, offering flexibility and adaptability that traditional systems lack.
    
    \item \textbf{Cost Efficiency}: By minimizing the reliance on third-party services and streamlining the process through smart contracts, our protocol reduces both the development and usage costs, making digital wills more accessible to a broader audience.

    \item \textbf{Funds remain at Rest}: By focusing on interchain messages, will creators are not required to move their funds from their host chains. Users can also use their traditional wallets. Typically digital asset inheritance plans require users to transfer their assets to a new address.
    
    \item \textbf{Interchain and Multi-asset}: Typically, traditional service providers mainly support bitcoin, or a very small amount of supported assets. By constructing a formal interchain protocol to serve as a foundation, we can help scale service providers offering interchain, multi-asset inheritance at the protocol \& smart contract layer

	\item \textbf{Traditional Providers can become Validators}: Typically, with digital innovation, the entities being disrupted are purposefully left out of the innovation. However, our protocol enables traditional providers to modernized themselves, and become incentivized to further the network, and execute wills.
	
\end{itemize}

\section{Types of Wills}

In the following sections, we examine the various implementations of a Will on Willchain, each with its own unique functionalities and benefits. We denote a will by, \(\texttt{W}\).

\subsection{Will Module}
We define a module dedicated to providing the will interface, and capabilities.

\subsection{Will Components}
Will consist of several will components, denoted \(\texttt{W}_c\). Clearly, a will must have at least one component.

\begin{equation}
|\texttt{W}_c| \geq 1
\end{equation}

\subsection{Component Types}

\begin{equation}
\texttt{W}_c := C_{transfer} \lor C_{claim} \lor C_{call} \lor C_{ibc}
\end{equation}

\subsubsection{Execution Component}
Execution components are designed to run upon will expiration. 

\subsubsection{Claim Component}
Claim component are designed to both run when the will is expiration. They can also be triggered to run after a specific time duration is reached once a beneficiary has submitted a claim. This time duration can be canceled if the will creator checks in during the time duration triggered by the beneficiary claim. In summary here, claim components can be designed to only be claimable after will expiration, or during if the will creator fails to checkin after the claim component has been triggered. This allows for flexibility in how claim components function.

\subsubsection{Contract Component}
Will creators can direct a will component to execute smart contract logic at execution time, and during claim time. This means will creators can instruct their will to invoke a smart contract when it expires, and also allow for subsequence claims to be submitted by their beneficiaries after the will's expiration.


\subsection{Basic Will as a Contract}

Although the current implementation of this specification is a layer-1 blockchain itself, the original proof-of-concept was implemented entirely using smart contracts, and off chain executors. For that reason, we will cover the original implementation of the most basic Will contracts, and the associated infrastructure needed -- this includes the contracts, off-chain zk circuit compilation, and the relayers (previously referred to as executors in this paper) needed to invoke the will contracts. When implementing wills as contracts, the off-chain executors/relayers become key to ensure the self-executing property --even though it wouldn't be self executing, as off-chain entities are invoking execution of the wills in this case.

The Basic Will contract is the simplest form of a digital will. It's a contract that verifies a signature, $s_{ \texttt{prkey}_b}$, from a beneficiary's private key, $\texttt{prkey}_{b}$. The beneficiary can claim the will's assets by submitting a valid signature.

\subsubsection{Verification Logic for contract}

The contract simply compares the result of \(\texttt{ecrecover}\) to the address specified by the original creator of the will contract. The signature is generated from a \(\texttt{sign}\) function $\varphi()$.

\begin{align}
    \texttt{prkey}_b = \varphi( {prkey}_b )
\end{align}

While this implementation worked, we noticed a few shortcomings. The first issue was the reliance of off-chain executors just to get any seemingly automatic execution. The second issue was with the limitation of the on-chain verification logic. 

\subsubsection{ZK Verification in Contracts}
By using open source frameworks to construct zero knowledge circuits, provers, and verifiers. We were able to create a verification contract in solidity. This gave us the capabilities to allow for custom ZK circuits, and an endless amount of ways to create privacy-preserved verification logic on-chain for private fields. These early proof-of-concepts led us to implement the same type of capabilities, at the protocol level so we can control which logic we put on-chain and off-chain.

\subsubsection{Token ID Creation}

At will creation time, a soulbound NFT with a unique token ID is minted to represent the will, denoted as $ID_{t_{dw}}$. The token ID is generated using the hash function $\varphi()$. The unique token ID is a decentralized identifier (DID).

\begin{align}
    ID_{t_{dw}} = \varphi(k_{ow}) \quad \text{where $k_{ow}$}
\end{align}

\subsubsection{Decentralized Identifiers}
We make use of DIDs by registering a formal DID method with the W3C.

The format introduces a \(\texttt{will}\) method, followed by the will's identifier, \(\texttt{W}_{id}\)

\begin{equation}
\texttt{did}:\texttt{will}:\texttt{W}_{id}
\end{equation}



\section{Mathematical Model of a Will}

Given the complexity of the digital will provided, we establish a comprehensive model that captures the nuances of each component, including the specific types of claims, their cryptographic mechanisms, and the outputs upon execution.

\subsection{General Representation of Will Components}

Each component \(C_i\) in the will \(W\) is represented by a tuple \((T, A, O, \delta)\), where:
\begin{itemize}
    \item \(T\) denotes the type (e.g., transfer, schnorr-claim, pedersen-claim, contract call, IBC message).
    \item \(A\) specifies access controls detailing the permissions required to activate or execute the component.
    \item \(O\) encapsulates the outputs that the component generates upon execution.
    \item \(\delta\) is the state transition function describing how component states change under certain conditions.
\end{itemize}

\subsection{Specific Components and Their Formulations}

\subsubsection{Transfer Component with Emission}

This component involves transferring a digital asset and emitting a message upon successful execution:
\begin{align*}
T &= \text{"transfer+emit"} \\
A &= \text{"private"} \\
O &= \text{"transferred\_the\_tokens", "to": "address3", "amount": 987654321, "denom": "uwill"} \\
\delta(s, x) &= 
\begin{cases} 
\text{"executed"} & \text{if valid transfer conditions are met} \\
\text{"inactive"} & \text{otherwise}
\end{cases}
\end{align*}

\subsubsection{Schnorr Claim with Transfer}

This component utilizes a Schnorr signature for validation and executes a transfer upon successful claim verification:
\begin{align*}
T &= \text{"schnorr-claim+transfer"} \\
A &= \{\text{"private", ["address1", "address2", "address3"]}\} \\
O &= \{\text{"transfer\_to": "address4", "amount": 1000000000, "denom": "uwill"}\} \\
\delta(s, x) &= 
\begin{cases} 
\text{"active"} & \text{if Schnorr verification is successful} \\
\text{"inactive"} & \text{otherwise}
\end{cases}
\end{align*}

\subsubsection{Pedersen Claim with IBC Send}

Incorporating a Pedersen commitment, this component sends assets over an IBC channel upon certain triggers:
\begin{align*}
T &= \text{"pedersen-claim+ibc-send"} \\
A &= \{\text{"private", ["address1"]}\} \\
O &= \{\text{"IBC\_channel": "channel-0", "IBC\_address": "uwill", "amount": 123, "denom": "uwill"}\} \\
\delta(s, x) &= 
\begin{cases} 
\text{"active"} & \text{if Pedersen commitment is valid} \\
\text{"inactive"} & \text{otherwise}
\end{cases}
\end{align*}

\subsubsection{Gnark Claim with Contract Call}

Utilizing zero-knowledge proofs (Gnark), this component calls a contract upon verification:
\begin{align*}
T &= \text{"gnark-claim+contract-call"} \\
A &= \{\text{"private", ["address1"]}\} \\
O &= \{\text{"contract\_address": "0xcontract\_address", "payload": "data"}\} \\
\delta(s, x) &= 
\begin{cases} 
\text{"active"} & \text{if Gnark proof is valid} \\
\text{"inactive"} & \text{otherwise}
\end{cases}
\end{align*}

\subsubsection{IBC Message with Emission}

This component sends a message through IBC and emits a notification:
\begin{align*}
T &= \text{"ibc-msg+emit"} \\
A &= \text{"public"} \\
O &= \{\text{"message": "sent ibc message"}\} \\
\delta(s, x) &= 
\begin{cases} 
\text{"executed"} & \text{if IBC message is sent successfully} \\
\text{"inactive"} & \text{otherwise}
\end{cases}
\end{align*}

\subsection{Will Execution and State Transition}

The will \(W\) contains an ordered set of components \(\{C_1, C_2, \ldots, C_n\}\) which are executed based on the specified conditions and their access controls. The overall state transition of the will can be expressed as:
\begin{equation}
S' = \bigcup_{i=1}^n \delta(s_{C_i}, x_i)
\end{equation}

\textbf{Execution Condition:} The will is triggered either by a specific time or event, leading to the sequential or conditional execution of its components, depending on their individual \(\delta\) functions.

\textbf{Output upon Execution:} Each component outputs according to its type and function, potentially altering states within the blockchain or external systems.

This model provides a robust framework for analyzing and implementing complex digital wills, ensuring that all conditions, actions, and transitions are clearly defined and effectively managed.



\section{Protocol Handshake}

\subsection{Contract Approval}
Before will creation, the will creator must approve the relevent contracts existing on the chains where their assets are located. 

\subsection{Will Creation}
Accounts can broadcast a will creation transaction to the chain, which will contain the data of the will. The will data consist of the will, expiration, beneficiary, and components. The protocol generates a unique decentralized identifier (DID) for the will. Each component also receives a unique component ID as well. At will creation time, an interchain soulbound NFT, following ICS 721, is minted to the will creator. 

\subsection{Will Expiration}
On chain, the will expiration is defined by a block height number. At each block produced, the protocol retrieves all Wills designated for expiration at the current block height. 

\subsection{Will Execution}
During the will expiration block, any native execution components contained in the will are automatically executed. These components can be to transfer the native token of willchain, or to invoke a smart contract, or to send an IBC message.

\subsubsection{Execution-time logic}
Execution-time logic represents logic that is automatically executed when the will expires. 

\subsubsection{Interchain Execution}
Upon will execution, the protocol publishes a message to the respective smart contracts deployed on destination chains. This allows for subsequent transactions to be sent to those contracts for beneficiary claims. 

\subsection{Beneficiary Claims}

\subsubsection{Early Claim Penalty}
If a beneficiary submits a claim transaction, and the will is not expired, the beneficiary is penalized by native tokens being burned from their account. This enforced desired behavior by beneficiaries. We enforce penalties by utilizing the interchain \(\texttt{init-ack-confirm}\) handshake that allows for will eligibility verification to occur during the execution of a interchain claim.

\subsection{Entrypoint Contracts}
At the time of writing v2, each smart contract enabled chain supported by the Willchain protocol has required a custom contract to be deployed onto the chain. We have defined a set of template contracts in both Solidity, and Rust to serve as the common specification for our entrypoint contracts. 

Each contract fundamentally exposes the capability to be reached via interchain message. Additionally, the contracts are approved by the will creator at will creation time, \(\texttt{W}_creation_t\)


\section{Claim Types}
In the following sections, we discuss the different methods a beneficiary can use to claim from a Digital Will.

\subsection{Direct Claim}

In a direct claim, the beneficiary's private key $k_{pb}$ is utilized to generate a signature $s$. This signature is then checked against the stored hashes in the will contract. 

\begin{align}
    s = \sigma(k_{pb}) \quad 
\end{align}

where $\sigma$ is the signature function. In the above equation, $k_{pb}$ represents the private key of the beneficiary. The symbol $\sigma$ denotes the function that generates the signature.

\subsection{Claim with Private Key Proof}

This claim method involves the submission of a proof $\pi_{pb}$ by the beneficiary. This proof contains their private key as a private field of the verification circuit. The smart contract then verifies this proof and releases the will's assets if the proof is found to be valid.

\begin{align}
    \texttt{validity} = \phi(\pi_{pb}, k_{pb}) \quad 
\end{align}

where $\phi$ is the verification function. In this context, $\pi_{pb}$ symbolizes the proof that includes the private key. The function $\phi$ is the verification function that checks the validity of the proof.

\subsection{Claim with Signature Proof}

This method involves the beneficiary submitting a proof of signature $\pi_s$. This proof is then verified against the stored hash of the signature $h_s$ in the will contract. Upon successful verification, the will's assets are released.

\begin{align}
    \texttt{validity} = \phi(\pi_s, h_s) \quad
\end{align}

where $\phi$ is the verification function. Here, $\pi_s$ represents the proof of the signature and $h_s$ is the hash of the stored signature. The function $\phi$ serves as the verification function.

\section{Types of Proofs}

\subsection{Proof Using Private Key as a Private Field}

This proof involves using the private key $k_{pb}$ as a private field in the verification circuit. The beneficiary submits the proof $\pi_{pb}$, and the smart contract checks its validity.

\begin{align}
    \texttt{validity} = \phi(\pi_{pb}, k_{pb}) \quad
\end{align}

where $\pi_{pb}$ represents the proof including the private key and the function $\phi$ is used for verification of the proof.

\subsection{Proofs Using Signatures}

In this type of proof, the beneficiary provides a proof of signature $\pi_s$. The smart contract verifies this proof against the stored hash of the signature $h_s$ in the will contract.

\begin{align}
    \texttt{validity} = \phi(\pi_s, h_s) \quad
\end{align}

In thiss scenario, $\pi_s$ symbolizes the proof of the signature while $h_s$ is the stored hash of the signature. The function $\phi$ checks the proof's validity.

\subsection{Proofs Using Homomorphic Encryption}

Pedersen commitments \cite{pedersen} enable beneficiaries to prove their knowledge of a certain piece of information (like a private key $k_{pb}$ or a signature $s$) without revealing that information. The smart contract checks the validity of the Pedersen commitment $P$ and releases the will's assets if the proof is valid. 

In the case of proofs using the homomorphic properties of Pedersen commitments, the beneficiary generates a Pedersen commitment $P$ using their private key $k_{pb}$ and a signature $s$ with the Pedersen commitment function $\psi$, so we have $P = \psi(k_{pb}, s)$.

Then, the validity of the commitment is checked using the verification function $\phi$. This function takes the Pedersen commitment $P$ as input and outputs a binary decision about the validity of the commitment. The equation can be expressed as: $\texttt{validity} = \phi(P)$.

Here, $P$ symbolizes the Pedersen commitment, $\psi$ stands for the Pedersen commitment function, $k_{pb}$ represents the beneficiary's private key, $s$ is the signature, and $\phi$ is the verification function.

\begin{align}
    P = \psi(k_{pb}, s) \\
    \texttt{validity} = \phi(P)
\end{align}

In this process, the Pedersen commitment provides a proof without revealing the private key or the signature, ensuring the privacy of the beneficiary's information. If the proof is valid, the assets of the will are released to the beneficiary.

\subsubsection{Pedersen Commitments and Homomorphism}

Pedersen commitments are cryptographic primitives extensively used in zero-knowledge proofs for their desirable homomorphic properties and perfect hiding properties. They are based on the hardness of the Discrete Logarithm Problem (DLP) in a cyclic group.

Let $G$ be a cyclic group of prime order $q$ with generator $g$. We introduce another group element $h = g^a$, where $a$ is chosen at random from $[1, q-1]$. In practice, $a$ should be unknown to ensure the security of the commitment scheme. For a message $m$ and randomness $r$, both chosen from $[1, q-1]$, a Pedersen commitment is constructed as follows:

\begin{align}
    P = \psi(m, r) = g^m h^r
\end{align}

One of the principal advantages of Pedersen commitments is their homomorphic property. Specifically, Pedersen commitments are both additively and multiplicatively homomorphic. Given two commitments $P_1 = \psi(m_1, r_1) = g^{m_1} h^{r_1}$ and $P_2 = \psi(m_2, r_2) = g^{m_2} h^{r_2}$, we can create new commitments by multiplying $P_1$ and $P_2$:

\begin{align}
    P_1 \cdot P_2 = g^{m_1+m_2} h^{r_1+r_2} = \psi(m_1+m_2, r_1+r_2)
\end{align}

This is an example of the additive homomorphism of Pedersen commitments, as we can create a commitment to the sum of two messages using the commitments to the individual messages.

These properties make Pedersen commitments particularly suitable for privacy-preserving protocols such as zero-knowledge proofs.

\subsection{Proofs using Schnorr Signatures}
In order to further enable privacy, we the protocol supports schnorr signatures \cite{schnorr} as a claim type. This allows wills to require signatures that can be constructed by aggregating \(n\) signatures together to create a single aggregated signature that can be used in verification. This scheme hides the amount of signatures used in the aggregation. This approach is also used in Bitcoin as of the Taproot upgrade.

%
%
%
%
%

\subsection{Schnorr Signature Aggregation}

Schnorr signatures are renowned for their simplicity and efficiency, providing both security and privacy in digital signature schemes. They leverage the properties of elliptic curve cryptography (ECC) to generate compact signatures.

\subsubsection{Schnorr Aggregate Signatures}

Schnorr signatures enable a novel form of signature aggregation, which significantly enhances privacy and scalability. In the context of digital wills, this capability allows multiple parties to jointly create a single signature that validates a transaction or claim without revealing individual signers' identities or the exact number of participants.

Consider a scenario where multiple beneficiaries are entitled to claim from a digital will under a common condition. Instead of individually submitting proofs, these beneficiaries can collaboratively generate a single aggregate Schnorr signature $S_{agg}$. This aggregated signature, representing a unanimous agreement or claim, maintains the privacy of individual participants by not disclosing their number or identities.

The mathematical foundation for generating an aggregate Schnorr signature involves each participant creating their own signature $(s_i, R_i)$ for a given message $m$, where $s_i$ is the signature component and $R_i$ is the nonce. The aggregate signature $S_{agg}$ is then constructed by summing the individual signatures and nonces:

\begin{align}
    S_{agg} = (\sum s_i, \sum R_i) \quad \text{mod} \quad q
\end{align}

where $q$ is the order of the group. The verification process for $S_{agg}$ uses the aggregated nonce $\sum R_i$ and the public keys of all participants, ensuring the signature is valid if and only if all individual signatures are valid.

\subsubsection{Privacy and Scalability Advantages}

Schnorr aggregate signatures offer notable privacy benefits by masking the number of signers involved, making it impossible to deduce individual participants in a multi-signature setup. This characteristic is particularly advantageous for digital wills, where preserving the anonymity of beneficiaries is desired.

Moreover, the aggregation reduces the blockchain's storage and bandwidth requirements, as a single aggregate signature occupies significantly less space than multiple individual signatures. This scalability improvement is crucial for complex wills involving numerous beneficiaries, ensuring the system remains efficient regardless of the number of participants.

\subsubsection{Security Considerations}

Schnorr signatures adhere to the highest security standards, relying on the intractability of the Discrete Logarithm Problem (DLP) in elliptic curves. The security model ensures that, even when signatures are aggregated, the resulting signature is as secure as individual Schnorr signatures, provided the cryptographic assumptions hold.

The inclusion of Schnorr aggregate signatures as a claim type in our protocol not only enhances privacy and scalability but also ensures robust security. The mathematical elegance and efficiency of Schnorr signatures make them an ideal choice for modern cryptographic applications, including the innovative realm of digital wills.



\subsection{Proofs using zk-SNARKs}

Let $k_{\text{priv}}$ be the private key and $x$ be the public inputs. A statement in zk-SNARKs can be represented as a function $\Lambda(k_{\text{priv}},x)$.

\subsubsection{Efficiency and Security of zk-SNARKs}
Zk-SNARKs are efficient due to their succinctness and quick verification time. However, one caveat is that zk-SNARKs require a trusted setup, which can be a security concern if not properly implemented.

For instance, let's consider a statement to be proved: "I know a secret number $w$, such that $h = g^w$, where $g$ and $h$ are public numbers." The prover can generate a proof $\pi = \Pi(k_{\text{priv}}, h)$ and the verifier can check this proof with $\beta(vk, h, \pi)$. It offers an unparalleled privacy standard, though, further research into optimizing the trusted setup phase can fortify this technology's security.

\subsubsection{Creating a proof}

Let $\Pi$ be the function that takes the private key $k_{\text{priv}}$ and public inputs $x$ and returns a proof $\pi$:

\begin{equation}
\pi = \Pi(k_{\text{priv}}, x)
\end{equation}

\subsubsection{Verifying a proof}

Let $\beta$ be the function that verifies a proof. The verification process is then represented by the function $\beta(vk, x, \pi)$, where $vk$ is the verification key and $\pi$ is the proof. The function returns a boolean value:

\begin{equation}
\beta(vk, x, \pi) = 
\begin{cases} 
      \text{true} & \text{if } \Lambda(x,k_{\text{priv}}) = 0 \\
      \text{false} & \text{otherwise} 
\end{cases}
\end{equation}

We intend to continue to add more will claim types as the community deems suitable.

\section{Checking In}

Checking into the will occurs natively on the chain by default. Let $\Theta(u, t)$ be the check-in function. The check-in process can be represented by the function $\Theta(u, t)$, where $u$ is a user and $t$ is the current time. This function returns a boolean value:

\begin{equation}
\Theta(u, t) = 
\begin{cases} 
      \text{true} & \text{if } u \text{ checks in at time } \leq t \\
      \text{false} & \text{otherwise} 
\end{cases}
\end{equation}

\subsection{Checkin Types}

\subsubsection{Native Checkin Transaction}
Accounts can submit a raw transaction to checkin, as the protocol exposes a native checkin function.

\subsubsection{Checkin from a Contract}
Developers can also write smart contracts that invokes the native checkin functionality offered by the will module. This allows for custom checkin logic. 

\subsection{Optimizing the Check-in Process}
The check-in function $\Theta(u, t)$ has significant implications on the system's reliability and the user's experience. An optimal check-in frequency should balance between the user convenience and the need for timely activation of the digital will. 

By allowing for checkins via transaction, or a smart contract, this allow for users to experiment with different security models, and conditions. Moreover, the check-in system should incorporate robust security measures to prevent unauthorized check-ins. By default, if an account checks in from a transaction, the will rejects any transaction from any other address other than the creator of the will. This is a built in feature to how transaction work in blockchains, and we are merely adhering to its common pattern. An interesting direction for research could be incorporating multi-factor authentication techniques into the $\Theta(u, t)$ function, making the check-in process more secure while still maintaining user convenience. Account abstraction also enables a more user friendly checkin process.


\section{Refungible Tokens for Fractionalization}

Optionally, destination contracts, and contracts deployed on our protocol can issue, and maintain temporary representative utility tokens to their will's beneficiaries. At will expiration time, this allows beneficiaries to use these utility tokens to make claims against the will, and receive will distribution directly correlated to their balance of representative utility tokens. Upon usage, these representative utility tokens can be burned or locked forever. This approach can also be done in smart contracts on Willchain, to allow for custom logic for fractionalizing beneficiary distributions.

Let's denote the total number of shares as $S$, and the individual share of heir $i$ as $s_i$. Then the sum of all shares equals the total number of shares:

\begin{equation}
S = \sum_{i=1}^{n} s_i
\end{equation}

The percentage share of each heir can then be calculated as follows:

\begin{equation}
p_i = \frac{s_i}{S}
\end{equation}

Where $p_i$ is the percentage of the total asset that heir $i$ will receive.

In the context of RFT logic, we define the equation for the calculation of shares:

\begin{equation}
S_{ij} = \Lambda_j(x_i)
\end{equation}

\subsection{RFT issued by a Will Contract}

A Refungible Token, or RFT, is a contract that owns an NFT, and also issues a fungible token that represents fractionalized ownership in the NFT. The RFT version of the Will contract extends the basic version by representing the will as a non-fungible token (NFT). The will's assets are tied to the token, denoted as $\texttt{W}_{token}$, and the beneficiary can claim them by submitting the correct inputs to the contract. Will creators can also write custom logic on how to distribute and handle shares of $\texttt{W}_{token}$.

\subsection{Efficiency and Fairness of an RFT}
An RFT ensures that digital assets are distributed fairly among the heirs, based on predefined, or runtime computed shares. Its efficiency lies in its direct computation and equitable distribution. Still, comparing it to traditional asset division methods might reveal its superiority or areas that need improvement. 

One area of investigation might be the conditional execution of RFT logic. For example, how should the system respond if a heir $i$ predeceases the user? Further mathematical modeling would be required in this context, refining the $S_{ij} = \Lambda_j(x_i)$ equation to account for such scenarios. The fairness becomes about how these tokens issued by the RFT are distributed amongst beneficiaries -- however, this too is programmable already.

\subsubsection{Claiming Assets}

In a pure contract implementation with NFT wills, the beneficiary claims assets by invoking the NFT transfer function $f_{tr}(a_{b}, ID_{t_{dw}})$, where $a_{b}$ is the Beneficiary Address. This is done only if the hash of the provided private key matches the stored hash.

\begin{align}
    h_{pb} = \varphi(k_{pb}) \\
    f_{tr}(a_{b}, ID_{t_{dw}})
\end{align}

\subsection{NFT Bound Accounts}

The EIP-6551 version of the DigitalWill extends the NFT/RFT version by representing the will as a token-bound account. This account, denoted as $A_{eip}$, can own other NFTs and operate on the blockchain on behalf of the token it's bound to. The beneficiary can claim the will's assets and the token-bound account by submitting the correct private key.

\subsubsection{Token-Bound Account Creation}

A token-bound account is created using the EIP-6551 standard. The creation function $\chi()$ takes several parameters, including the chainId $c$, tokenContract $tc$, tokenId $tid$, and a salt value $s$.

\begin{align}
    A_{eip} = \chi(\texttt{impl}, c, tc, tid, s)
\end{align}

\subsubsection{Claiming Assets}

The beneficiary claims assets and the token-bound account by providing the correct signature and invoking the NFT transfer function. In the process of claiming assets, two primary operations take place. First, the signature is verified by the contract. Here, $s_{ \texttt{prkey}_b}$ represents the signature from $\texttt{prkey}_b$. This is the same setup as the aforementioned basic will contract example.

Subsequently, the function $f_{tr}(a_{be}, \tau)$ is invoked to effect the transfer of the will NFT. In this function, $f_{tr}$ represents the transfer function, $a_{be}$ is the beneficiary's address, and $\tau$ stands for the unique ID of the will NFT.

\begin{align}
    h_{pb} = \varphi(k_{pb}) \\
    f_{tr}(a_{be}, \tau) 
\end{align}


\section{Account Abstraction}

Let $\Phi(u, a)$ be the function that associates a user with an account. Account abstraction can be represented by the function $\Phi(u, a)$, where $u$ is a user and $a$ is an account:

\begin{equation}
\Phi(u) = a
\end{equation}

\subsection{Security and Privacy in Account Abstraction}
The function $\Phi(u, a)$ abstracts user accounts, providing a layer of privacy and security. However, it also presents potential security challenges that need careful consideration. 

How might $\Phi(u, a)$ protect against potential linking of a user to an account by malicious parties? Exploring different cryptographic techniques for enhancing privacy in account abstraction can be a potential research area. Also, considering multiple blockchains, how could $\Phi(u, a)$ evolve to ensure seamless interaction across chains?

Account abstraction would, at the most basic level, allow for users to have their checkin costs covered by third parties, including the foundation. There is much work to be done in the industry as a whole regarding implementing Account Abstraction in an optimal way.

\section{Modular Smart Contract Accounts}
We make use of a modern ERC specification that enables Modular Smart Contract Accounts, (MSCA), specified by ERC-6900. This enables accounts to not be governed by private keys, as traditional Externally Owned Accounts require, but to instead be controlled via smart contracts. This enables another level of confidence for asset holders. 

\subsection{Enhancing Will Execution MSCA}
The MSCA standard marks a revolutionary step in the development of smart contracts, especially in the realm of digital inheritance. This standard introduces modular smart contract accounts that significantly enhance the flexibility and interoperability of will execution mechanisms across various blockchain platforms.

\subsection{Foundation and Application to Digital Inheritance}

The MSCA framework provides a structured approach to developing smart contract accounts that can be tailored to specific needs, including those of digital wills. These accounts are constructed from discrete modules, each responsible for handling different aspects of the contract's functionality. This modular architecture is succinctly expressed as:

\begin{equation}
S_{MSCA}^{will} = \bigoplus_{i=1}^{n} W_i
\end{equation}

where $W_i$ denotes the $i^{th}$ module tailored specifically for the digital inheritance process within the MSCA, $S_{MSCA}^{will}$. The $\bigoplus$ symbol illustrates the modular composition, enabling a customizable and upgradable will execution process.

\subsection{Cross-Chain Will Execution}

A crucial advantage of employing the MSCA standard in digital wills is its inherent support for cross-chain operations. This capability allows for the seamless execution of wills across different blockchain networks, broadening the scope of asset distribution and simplifying the inheritance process for assets stored on diverse platforms.

\begin{equation}
\forall j, \exists T_{ij}^{will} : S_{MSCA}^{will} \rightarrow C_j
\end{equation}

This formula demonstrates that for any blockchain platform $C_j$, there is at least one transaction $T_{ij}^{will}$ that can be initiated by the MSCA-based will smart contract, ensuring comprehensive asset management across ecosystems.

\subsection{Advantages in Digital Will Creation and Management}

\subsubsection{Interchain Will Creation}

The MSCA standard promotes a unified method for creating and managing digital wills on various blockchains. This interoperability is paramount for a truly decentralized inheritance system, as it ensures that no asset, regardless of its native chain, is left behind in estate planning.

\subsubsection{Dynamic Will Modification}

The modularity of MSCA smart contracts allows for dynamic updates to the will's conditions or beneficiaries without the need to deploy a new contract. This adaptability is crucial for long-term estate planning, where changes in the testator's wishes or circumstances may occur.

\begin{equation}
\text{Flexibility} = f\left(\bigoplus_{i=1}^{n} \Delta W_i\right)
\end{equation}

Here, $\Delta W_i$ represents the change in the $i^{th}$ module of the will smart contract, with $f$ denoting the function that integrates these changes, emphasizing the flexibility of the MSCA standard. Prior to this, a typical approach to achieving a Dynamic will would be to implement the standard proxy pattern formalized by ERC-1822, for Universal Upgradeable Proxy Standard (UUPS). Usage of the protocol supports using both of these specs, with more preference place on ERC-1822.

\subsubsection{Universal and Secure Inheritance Protocol}

Leveraging MSCA for digital wills not only broadens the applicability of wills across chains but also enhances security through standardized, audited contract modules. This approach mitigates risks associated with custom contract development and ensures a high level of trust in the digital inheritance process.

\begin{equation}
\texttt{Security Assurance} = \int_{S_{MSCA}^{will}}^{} \sigma(ds)
\end{equation}

$\sigma(ds)$ quantifies the security assessment for each differential section $ds$ of the smart contract, highlighting the comprehensive security benefits of the MSCA standard in digital inheritance.

The integral notation here, $\int_{S_{MSCA}^{will}}^{} \sigma(ds)$, serves as a symbolic representation of the cumulative security assurance provided by the MSCA (Modular Smart Contract Account) framework, specifically tailored for digital wills ($S_{MSCA}^{will}$). In this context, $\sigma(ds)$ is a function that measures the security level or assessment for a differential segment ($ds$) of the smart contract system. Essentially, it evaluates how secure each part of the smart contract is against potential vulnerabilities or threats.

The use of integration in this formula is metaphorical, indicating a continuous, thorough examination across the entire structure of the MSCA. This meticulous approach ensures that every module or component of the MSCA contributes positively to the overall security of the digital will system. The integral symbolizes the aggregate security benefits derived from each segment of the contract, emphasizing that the MSCA framework's strength lies in its comprehensive coverage and the synergistic effect of its modular components.

In simpler terms, this formula suggests that by analyzing and securing each module of an MSCA individually, the overall security of the digital will system can be maximized. This integrated security assessment approach ensures that the digital will is robustly protected across all its aspects, from asset allocation to beneficiary designation and beyond.

\subsection{Forward-Looking: Towards a Multichain Inheritance Standard}

Our ongoing efforts focus on extending the MSCA specification to support additional smart contract languages and blockchain platforms, aiming to establish a universally accepted standard for digital wills. This endeavor seeks to provide a robust foundation for will creators, enabling them to confidently secure their digital legacy across the ever-expanding multichain landscape.

The adoption of the MSCA standard in digital inheritance systems introduces a new level of efficiency, security, and universality to the process of asset distribution. As blockchain technology continues to evolve, the principles laid down by ERC-6900 will undoubtedly play a central role in shaping the future of digital asset accounts.


\section{Executors}
When the Willchain project first began, and consisted of mainly smart contracts, we developed an off-chain executor to facilitate the desired flow of our protocol. As the protocol has matured into a layer-1 blockchain, we've implemented the Inter Blockchain Communication (IBC) protocol, whereby executors have evolved into IBC relayers. This has enabled the protocol to use an industry leading specification at the root of howe handle interchain communication.

Using IBC allows us to take advantage of the protocol handshakes that occur for channel management, and packet transmission. We make use of the \(\texttt{init-ack-confirm}\) information flow between IBC ports. The will module is always one end of the path being used. This ensures information is always coming from the will module, or going to the will module.

\section{IBC Relayers as Executors}

Let $\Psi(s, x)$ denote the state transition function induced by executing transactions or IBC packet processing. The IBC relayers can be conceptualized as the facilitators of cross-chain communication, enabling the function $\Psi(s, x)$, where $s$ is the current state of the blockchain and $x$ are the inputs from other chains:

\begin{equation}
s' = \Psi(s, x)
\end{equation}

IBC relayers form a set $R = \{R_1, R_2, ..., R_n\}$, where each $R_i$ represents an individual relayer responsible for monitoring state changes and relaying packets between chains. Let $\Omega_i$ correspond to the operation performed by relayer $R_i$:

\begin{equation}
\Omega_i(\chi, \rho) \rightarrow \zeta
\end{equation}

Each relayer employs a decision-making mechanism $\Delta_i$, determining the actions based on the proof of state changes and packet commitments:

\begin{equation}
\Delta_i(\pi, x) \rightarrow \texttt{Decision}
\end{equation}

Let $\Gamma$ represent the encapsulation of a contract's logic into an IBC packet. The transformation process is defined as:

\begin{equation}
\texttt{Contract}_{Logic} \xrightarrow{\Gamma(\cdot)} \chi
\end{equation}

The logic $\chi$ can be interpreted and acted upon by the destination chain's smart contract through the IBC application protocol.

Let $\Xi_i$ encapsulate the comprehensive decision-making and execution process managed by IBC relayers:

\begin{equation}
\Xi_i(\Delta_i(\pi, x)) \rightarrow \zeta
\end{equation}

\subsection{Reliability and Decentralization of IBC Relayers}
IBC relayers, denoted by $\Psi(s, x)$, are pivotal in ensuring the integrity and decentralization of cross-chain communication. They eliminate single points of failure, thus enhancing the system's resilience.

Critical considerations include optimizing the relayer set for efficiency without compromising decentralization and establishing protocols for handling misbehavior. 

The decision-making process, $\Delta_i(\pi, x)$, of relayers benefits from consensus mechanisms ensuring the validity of cross-chain transactions. 

\subsection{Security in IBC Relayers}
Security within the context of IBC relayers entails safeguarding the relay process and ensuring the integrity of the data being transmitted. A robust authentication and verification mechanism is paramount for detecting and mitigating misbehavior during the init-ack-confirm handshake phases of the IBC protocol.

Exploring advancements in cryptographic techniques to secure the packet relay process against manipulation and unauthorized access is essential. Additionally, implementing thorough misbehavior detection algorithms that can accurately identify and penalize malicious relayers could significantly enhance the security and reliability of cross-chain communications.


\section{Multisigs}
Although usage of multisigs doesn't allow for the same capabailities as a protocol, there is still much usefulness for multisigs in the protocol. Use of multisig accounts can be used for accounts on the main chain, accounts on destination chains, and combined with smart contracts can allow for any shared key setup.


\section{Revealing Keys to Beneficiaries}
If simple multsigs do not suffice for a use case, we support several methods of revealing keys. This can trivially be done by having a single key pair for the foundation, and privately send contents from the will to the beneficiary, whereby then they could verify its integrity against what is on chain. Let us try to do this in a way with multiple keys, to decrease the impact of the foundations keys becoming compromised. 

This method can be used if there is a need to reveal data to the beneficiary in a decentralized way. We can use the beneficiary's public key to encrypt a message. 

\begin{equation}
\lambda(\lambda(d, k_b), k_t)
\end{equation}

Where \(\lambda(d, k_b)\) is the encryption function on the data, \(d\), and using the beneficiaries key,  \(k_b\). \(\lambda(d, k_b)\) as input to the same encryption function, but using the temporary key, \(k_t\). We then use a temporary key to encrypt the result of the encryption via the beneficiary's key, \(k_b\), and use it as input in the encryption using the temporary key, \(k_t\). The temporary keys can be held by any party.

\begin{equation}
E_{k_b}(m) = c_1, \quad E_{k_t}(c_1) = c_2
\end{equation}

Here, $E_{k_x}(y)$ represents the encryption of message $y$ using key $k_x$, with $m$ being the original message containing sensitive data intended for the beneficiary. The key $k_b$ is the public key of the beneficiary, resulting in ciphertext $c_1$. Subsequently, $c_1$ is encrypted with a temporary key $k_t$, yielding the final ciphertext $c_2$. This layered encryption ensures that the message remains secure and only accessible under predetermined conditions.

\subsection{Strengths}

This method's primary advantages include enhanced security measures and procedural flexibility:
\begin{itemize}
    \item \textbf{Enhanced Security:} The initial encryption with the beneficiary's public key ensures that only the intended recipient with the corresponding private key can decrypt the message, thus safeguarding against unauthorized access.
    \item \textbf{Controlled Revelation:} Employing a temporary key introduces an element of control over when the encrypted data can be revealed, allowing the data to be securely distributed or stored until the conditions for key revelation are met.
    \item \textbf{Flexibility and Revocability:} The method allows for changes in the will or designated beneficiary, as the encrypted message can be revoked or altered as long as the temporary key remains undisclosed.
\end{itemize}

\subsection{Challenges with Revealing Temporary Keys}

Despite its strengths, this encryption method has several potential limitations:

\begin{itemize}
    \item \textbf{Key Management Dependency:} The method's efficacy and security depend on the robust management of the temporary key ($k_t$). A compromise of this key could lead to unauthorized access to the encrypted message.
    \item \textbf{Increased Complexity:} The need for additional keys and encryption steps may complicate the execution process of the will, increasing the potential for user error and necessitating a more sophisticated infrastructure for decryption.
    \item \textbf{Third-party Trust:} If a third party holds the temporary key, there is an implicit trust requirement. Mismanagement or malintent by this entity could compromise the security of the process.
\end{itemize}

Employing public key encryption augmented by a temporary key presents an effective strategy for key revelation in the context of digital wills. While offering notable security and control benefits, it also highlights the importance of addressing key management challenges and the complexity introduced by additional encryption layers. Trusted holder of the key, whether it be the foundation, or a third party, will have the same security requirements as a financial exchange. Robustness of the protocol would be a last line of defense against any nefarious behavior here. With proper penalties implemented, this approach, or an augmented version may provide practical key revelation. More research work is to be done on this. This method exemplifies the inherent trade-offs between enhancing security and managing procedural intricacy in digital inheritance planning, emphasizing the necessity for developing user-friendly, secure solutions.


\section{File Storage via Indexed Contracts}

Blockchains have traditionally faced challenges when tasked with storing large files due to inherent limitations in contract size. To address this, we introduce a novel, practical approach comprising two types of smart contracts: a mapping contract and storage contract(s). This design not only circumvents the size constraints but also ensures query efficiency and privacy preservation for encrypted data. This can obviously also be done at the protocol level, but we've chosen to limit it to the contract layer for the initial version.

\subsection{System Architecture}

The system is architecturally divided into two core components:

\begin{enumerate}
    \item \textbf{Mapping Contract:} Serves as an index that maintains a mapping of data identifiers to storage contract addresses and their respective indices, facilitating the location of file chunks across the blockchain.
    \item \textbf{Storage Contracts:} A collective of contracts that store pieces of the larger file, each operating within the size limits imposed by the blockchain platform.
\end{enumerate}

\subsection{Operational Workflow}

The operational workflow involves chunking large files, storing these chunks across multiple storage contracts, and indexing these locations within the mapping contract. Mathematically, let $F$ be a file of size $S$ bytes. We define a chunking function $\mathcal{C}$ that divides $F$ into $n$ chunks:

\[
\mathcal{C}(F) = \{F_1, F_2, \ldots, F_n\}, \quad \text{where } \sum_{i=1}^{n} |F_i| = S
\]

Each chunk $F_i$ is stored in a storage contract, indexed by the mapping contract. Let $M$ denote the mapping contract, and $\mathcal{SC} = \{SC_1, SC_2, \ldots, SC_m\}$ represent the set of storage contracts. The mapping stored in $M$ can be represented as:

\[
M(F_i) = (SC_j, idx), \quad \forall F_i \in \mathcal{C}(F)
\]

where $SC_j$ is the storage contract address and $idx$ is the index within $SC_j$ pointing to $F_i$.

\subsection{Querying for Data}

To reconstruct the file $F$, a user queries $M$ to obtain the mapping of chunks, and subsequently queries the corresponding storage contracts. The reconstruction function $\mathcal{R}$ combines the chunks retrieved from $\mathcal{SC}$:

\[
\mathcal{R}(\{F_1, F_2, \ldots, F_n\}) = F
\]

\subsection{Privacy-Preserving Encrypted Data Storage}

Encrypted data storage introduces an additional layer of privacy. Let $\mathcal{E}$ be an encryption function where $\mathcal{E}(F) = F'$, and $F'$ is the encrypted form of $F$. The chunking and storage process follows as previously described but with $F'$ instead of $F$.

This methodology becomes particularly advantageous for will execution. Encrypted legal documents, such as property deeds, can be securely stored on-chain. Upon the grantor's demise, the decryption key is released to the beneficiary, enabling them to access and reconstruct the legally binding document.

\subsection{Example: Property Deed Transfer}

Consider a will creator intending to transfer a property deed. The document is created, notarized, and encrypted resulting in $F'$. Following our proposed architecture, $F'$ is chunked and stored across blockchain contracts. This ensures that the deed remains tamper-proof, confidential, and readily available for the beneficiary upon the release of the decryption key, thereby fulfilling the conditions for a legally binding transfer of property.

This approach leverages smart contracts to efficiently store large files on chain without compromising on privacy or legality. This approach holds particular promise for executing wills in a secure, transparent, and immutable manner, ensuring that sensitive files, like property deeds, are preserved and transferred according to the grantor's wishes.


\section{Concluding}

In conclusion, this proposal integrates advanced cryptographic primitives with an layer-1 blockchain that invokes interchain heterogenous smart contracts existing in multiple chains to create a secure, privacy-preserving, and self-executing digital will protocol. This framework embodies the potential of blockchain technology in reshaping legal and personal domains. As we continue to evolve and digitize, this platform demonstrate the transformative potential of blockchain technology, revolutionizing traditional systems and paradigms.  

\subsection{Active Areas of Research}
The protocol mentioned above was implemented for chains that are smart contract enabled, whether it be Rust or Solidity. This can also work on other chains with other contract languages, such as Cadence, Move, etc. However, a focus area of ours in the near future is to develop an IBC-compatible framework for Bitcoin. This is very early work, but the point would be to use features introduced by the Taproot Upgrade to formalize a handshake that can be made via an IBC relayer connected a path between Bitcoin and an IBC enabled blockchain.

\subsubsection{More Entrypoint Contract Types}
We are also continuously enhancing the entrypoint contracts, as new specifications are formalized, and proposed. It is essential to enable wills created by the protocol to support upgrades whenever possible. This paper serves as a brief summary of what we are focused on, and how we think through the problem and solution spaces. 

\subsubsection{Incentive Compatibility of Protocol}
Another proposal we are testing is the impact of different network properties towards aligning incentives of will creators, beneficiaries, validators, and relayers. One idea is to design an incentive which places bounties on will to be properly executed, as the protocol already penalizes early execution of a will. Experimentation has been done using threshold decryption, by splitting up a key amongst a set of actors. These actors can be partners, validators, and/or relayers.

\end{document}